\documentclass[aps,pra,twocolumn,epsfig,amsmath,floatfix]{revtex4}

\usepackage{xcolor}
\usepackage{epsfig,amsmath}
\usepackage{graphicx}
\usepackage{dcolumn}
\usepackage{bm}
\usepackage[]{amsmath}
\usepackage{braket}
\usepackage{amsfonts,amsmath}
\usepackage{epsfig,amsmath}
\usepackage{graphicx}
\usepackage{dcolumn}
\usepackage{bm}
\usepackage{bbold}

\usepackage{lineno}
\usepackage{xcolor}
\usepackage{epsfig,amsmath}
\usepackage{graphicx}
\usepackage{dcolumn}
\usepackage{bm}
\usepackage[]{amsmath}
\usepackage{braket}
\usepackage{amsfonts,amsmath}
\usepackage{epsfig,amsmath}
\usepackage{graphicx}
\usepackage{dcolumn}
\usepackage{bm}
\usepackage{bbold}
\newcommand{\beq}{\begin{equation}}
\newcommand{\eeq}{\end{equation}}
\newcommand{\beqa}{\begin{eqnarray}}
\newcommand{\eeqa}{\end{eqnarray}}

\newcommand{\la}{\langle} 
\newcommand{\ra}{\rangle}

\def\josaa#1{{ J. Opt.\ Soc.\ Am.\ A\/} {\bf#1}}

\def\jpa#1{{ J.\ Phys.\ A} {\bf#1}}

\def\natcom#1{{ Nat.\ Commun.} {\bf#1}}

\def\oe#1{{ Opt.\ Express} {\bf#1}}

\def\ol#1{{ Opt.\ Lett.} {\bf#1}}
\def\opt#1{{ Optica} {\bf#1}}

\def\pra#1{{ Phys.\ Rev. A\/} {\bf#1}}

\def\prx#1{{ Phys.\ Rev. X\/} {\bf#1}}
\def\prl#1{{ Phys.\ Rev.\ Lett.} {\bf#1}}
\def\prr#1{{ Phys.\ Rev.\ Research} {\bf#1}}

\def\sci#1{{ Science} {\bf#1}}
\def\rmp#1{{ Rev. \ Mod. \ Phys.} {\bf#1}}

\begin{document}

\title{Superresolution picks entanglement over coherence}
\author{Abdelali Sajia}
\author{Xiao-Feng Qian}
\email{xqian6@stevens.edu}
\affiliation{Center for Quantum Science and Engineering, and Department of Physics, Stevens Institute of Technology, Hoboken, New Jersey 07030, USA}

\begin{abstract}
Fundamental wave features of light play an important role in the realization of superresolution for two spatially separated point sources. It has been shown that (partial) coherence, which is an inevitable feature in practical light propagation, is harmful to measurement precision thus preventing superresolution. Here we study the quantitative effect of another fundamental feature, entanglement, on the quality of superresolution and compare it with that of coherence. Both single- and two-parameter estimations are analyzed in detail. Surprisingly, contrary to coherence, it is found that superresolution measurement precision (in terms of Fisher Information) can be enhanced as the amount of entanglement increases. More importantly, our analysis shows that non-zero entanglement always guarantees the non-vanishing of Fisher Information. Thus, while coherence is unwanted, entanglement is a favorable feature for superresolution.
\end{abstract}



\maketitle

\section{Introduction}

Entanglement and coherence are two fundamental features of a wave theory and are widely regarded as key resources in quantum information and quantum computation tasks \cite{NC, Plenio2017RMP}. Despite the fact that they originate from the same root (i.e., wave superposition) and are sometimes quantitatively connected \cite{Streltsov2015PRL}, the two features are fundamentally distinct physical concepts. Coherence, in most cases, can roughly be recognized as the capability to interfere, while entanglement defines the non-separability of two or more degrees of freedom (or vector spaces). Coherence will in general vary according to the observer's measurement basis \cite{Plenio2017RMP, BornWolf} while entanglement is independent of observation and is invariant under arbitrary local unitary basis changes \cite{horodoski-RMP}. More interestingly, in open quantum systems, coherence tends to vanish asymptotically \cite{zurek-decoherence} while entanglement can experience abrupt changes that lead to the phenomena of sudden death \cite{Yu2014PRL, Yu2017Sci}. Here instead of focusing directly on the properties of entanglement and coherence themselves, we extend the exploration of their distinctions by investigating their effects on a quantum task, i.e., superresolution of two-point sources.

The capability of resolving two separated point sources is long known to be restricted by the Abbe-Rayleigh diffraction criterion \cite{Abbe1873, Rayleigh1896, BornWolf,Rayleigh1979}. Recently, it has been demonstrated that this long-standing restriction can be overcome when the signal is analyzed through the phase-preserving spatial mode decomposition method instead of the traditional phase-insensitive direct intensity measurement \cite{Tsang2015O, Tsang2016PRL, Tsang2016PRX, Paur2016O, Rehacek2017PRA, Larson2018O, Hradil2019O, Liang2021O, Liang2023OE, Wadood2021OE}. Similar to traditional imaging and resolution, incoherence is an essential element to guarantee the success of this novel superresolution method by maintaining high measurement precision (i.e., the non-vanishing of Fisher Information) \cite{Tsang2015O, Tsang2016PRL}. It has been shown that the practically non-controllable \cite{Hradil2021OL} inevitable partial coherence can be very harmful to measurement precision and will result in the resurgence of the so-called Rayleigh's curse \cite{Larson2018O, Larson2019O}. On the other hand, recent results have shown that the inclusion of an entangled partner can enhance the quality of superresolution \cite{Sajia2022PRR}. This suggests that entanglement may be a favorable feature for superresolution. Several questions follow naturally. How exactly is entanglement affecting superresolution? Does the amount of entanglement matter? How does it compare to the effect of coherence?

To address these questions, here we investigate the superresolution of two point sources of which the spatial degree of freedom (DoF) is entangled with the field's remaining DoFs. The effect of the amount of entanglement (in terms of the quantitative measure concurrence \cite{Wootters}) is analyzed in detail. For comparison, we also examine the quantitative effect of degree of coherence in the exact same context. Consistent with previous literature \cite{Larson2018O}, it is found that coherence is indeed a harmful feature that destroys the measurement precision quantified by Fisher information \cite{Helstrom1976, Shukur2020Nat, LiuJPA2020}. Surprisingly, entanglement is shown to have the opposite effect. For the single-parameter estimation case (i.e., only the two-source separation is unknown), the increase of entanglement will significantly enhance Fisher information while the increase of coherence will destroy it. For the double-parameter estimation case, i.e., an additional parameter (either coherence or entanglement) is also unknown, it is discovered that Fisher information can be enhanced by any finite amount of entanglement but is insensitive to the value of coherence. Therefore, non-zero entanglement will guarantee a non-vanishing Fisher information thus permitting credible superresolution. Our result suggests that in order to have high-quality superresolution, it should not only have entanglement but also include as much amount as possible. These results reveal a new aspect of fundamental distinctions between coherence and entanglement on quantum measurement estimations.


\section{Two point sources}
We start with the consideration of an incoming field from two equal-intensity point sources separated by a distance $s$. At a shift-invariant imaging plane, the spatial domain of the two sources, denoted respectively as $h_{\pm}(x)=h(x\pm s/2)$, can be viewed as displacements from the amplitude function $h(x)$. The point-spread function (PSF) is taken to be in the Gaussian form $h^{2}(x)=\frac{1}{\sqrt{2\pi\sigma^{2}}}\exp[{-\frac{x^{2}}{2\sigma^2}}]$ with $\sigma$ being the width. Then the general normalized two-source field can be described as 
\begin{eqnarray}
\label{ent}
\ket{\Psi}=\frac{1}{\sqrt{2}}(\ket{h_{+}}\ket{\phi_1}+\ket{h_{-}}\ket{\phi_2}),
\end{eqnarray}
where $|h_{\pm}\ra$ are normalized vector representations of the spatial functions, $\braket{x|h_{\pm}}=h_{\pm}(x)$, and $|\phi_{1,2}\ra$ are two normalized amplitude functions of all remaining DoFs (e.g., temporal domain, polarization, etc.) for the corresponding sources. Here $\ket{\phi_1}$ and $\ket{\phi_2}$ are generic vectors with arbitrary overlaps, i.e., $|\gamma|=|\la \phi_1|\phi_2\ra| \in [0,1]$, usually referred to as the degree of coherence \cite{BornWolf}. When combined with the fact that $|h_{+}\ra$ and $|h_{-}\ra$ don't perfectly overlap (due to non-zero separation $s>0$), the above field represents an entangled state between the spatial DoF characterized by $\ket{h_{+}}$, $\ket{h_{-}}$ and the remaining DoFs described by $|\phi_{1}\ra$, $|\phi_{2}\ra$. 

For the convenience of following analysis, we re-express the remaining DoFs in a new basis, i.e., $\{\ket{\phi_1},\ket{\phi_1^{\perp}}\}$, where $\ket{\phi_2}=e^{i\varphi}\cos\theta\ket{\phi_1}+\sin\theta\ket{\phi_1^{\perp}}$ and $\la \phi_1|\phi_1^{\perp}\ra=0$. Then the optical field (\ref{ent}) can be rewritten as
\begin{eqnarray}
\label{estimatedstate}
\ket{\Psi}=\ket{\Phi_{1}}\ket{\phi_1}+\ket{\Phi_{2}}\ket{\phi_1^{\perp}},
\end{eqnarray}
where $\ket{\Phi_{1}}=(\ket{h_{+}}+e^{i\varphi}\cos\theta\ket{h_{-}})/\sqrt{2}$ and $\ket{\Phi_{2}}=e^{i\varphi}\sin\theta\ket{h_{-}}/\sqrt{2}$ are two non-normalized spatial amplitudes. 

Now one can immediately achieve $\gamma=e^{i\varphi}\cos\theta$, which yields the degree of coherence as $|\gamma|=\cos\theta$. On the other hand the degree of entanglement, in terms of concurrence \cite{Wootters}, of the field can also be straightforwardly computed to be $C=|\sin\theta|\sqrt{1-d^2}$, with $d=\exp[-s^{2}/8\sigma^{2}]$.


\section{One parameter estimation}
The measurement of the target parameter, i.e., the two-source separation $s$, is always compensated with an indefinite variance. To characterize the measurement precision, we adopt Fisher information (FI), a popular quantity in estimation theory \cite{Helstrom1976, Shukur2020Nat} representing the average information per photon that can be extracted about an unknown parameter from a measurement. In this section, we consider the case where the separation $s$ is the only unknown parameter. We treat the measurement as a single repetition and do not consider the environmental loss cases where the photon numbers are also unknown.

From the Cram\'er-Rao theory \cite{Kay1993, Zmuidzinas2003OSA} Fisher information $F$ corresponds  to the lower-bound of the variance of the estimator $s$, i.e., $Var(s)\geq 1/F$. It can be specifically computed as $F_{\rho}=2{\rm Tr}[(\partial_s\rho(s))^2]$ for an arbitrary pure state, e.g., $\rho (s)=|h\ra\la h|$. For the two-source field (\ref{estimatedstate}), the spatial components are in a mixed state due to the remaining DoFs characterized by $|\phi_1\ra$, $\ket{\phi_1^{\perp}}$. Therefore, we adopt a weighted Fisher information \cite{Hradil2019O,Sajia2022PRR} based on the two possible measurement outcomes, i.e., $F_{tot}=\braket{\Phi_{1} |\Phi_{1}}F_{1}(s)+\braket{\Phi_{2} |\Phi_{2}}F_{2}(s)$, where $F_{1,2}(s)$ correspond to the Fisher information of the state $|\Phi_{1,2}\ra$ respectively. Straightforward calculation leads to the weighted FI as, 
\beqa
\label{FtotC}
F_{tot}&=&\frac{1}{4\sigma^{2}}-\frac{d(4\sigma^{2}-s^{2})\sqrt{1-d^2-C^2} }{16\sqrt{1-d^2}\sigma^{4}} \notag \\
&&-\frac{d^{2}s^{2}(1-d^2-C^2)}{8(2-2d^2-C^2+2d\sqrt{1-d^2}\sqrt{1-d^2-C^2})},
\eeqa
which is a function of the separation $s$ and entanglement $C$. This is our first major result. 

\begin{figure}[t!]
\includegraphics[width=6.5cm]{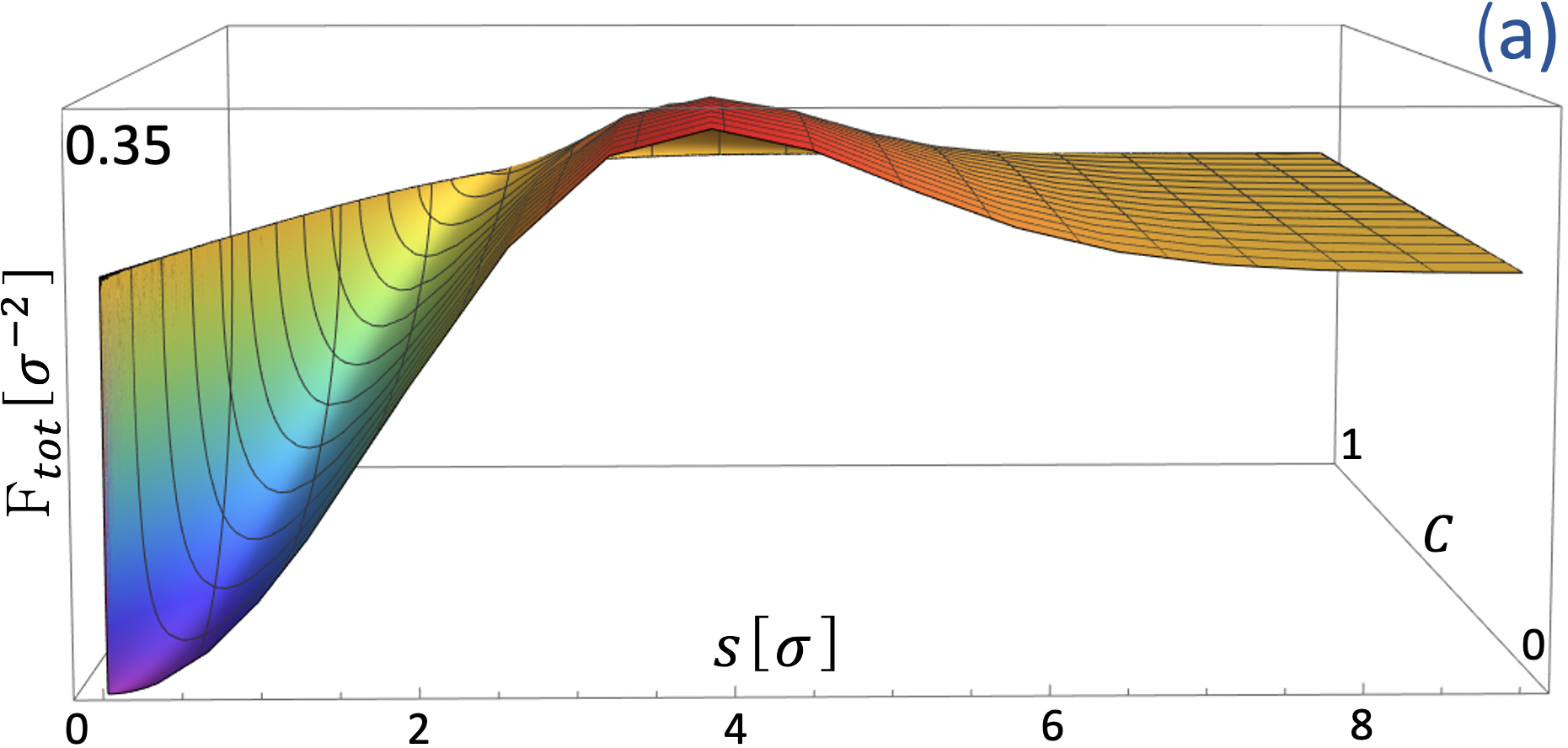}
\includegraphics[width=6.5cm]{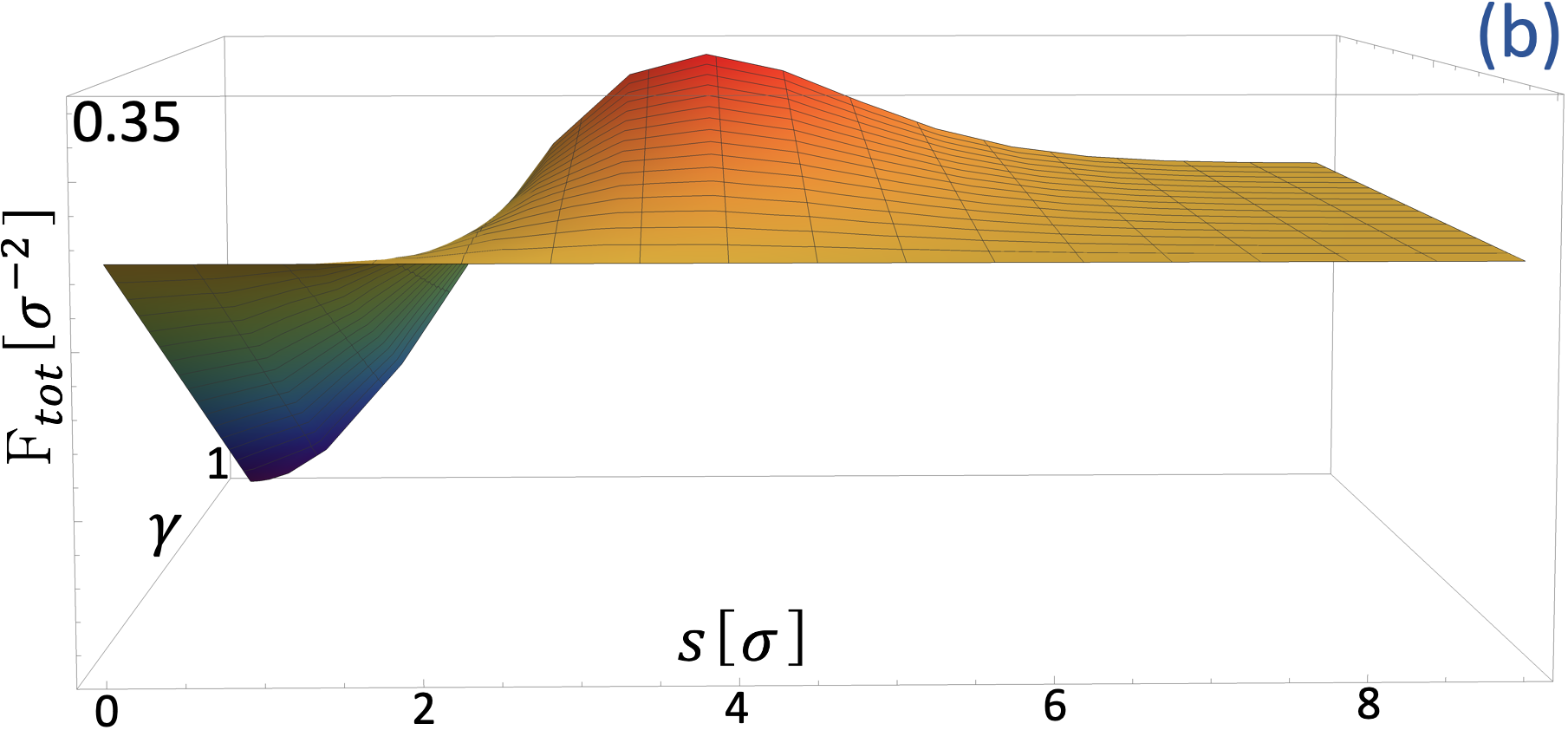}
\includegraphics[width=6.5cm]{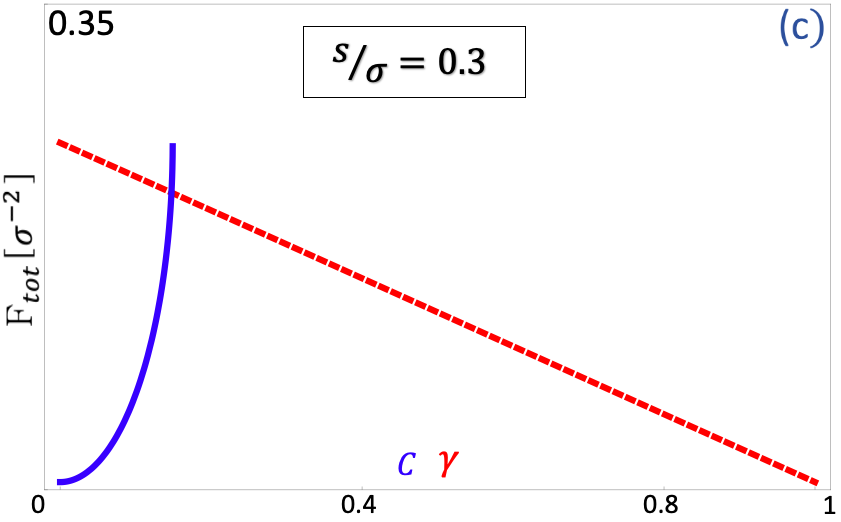}
\caption{Total Fisher information for $\sigma=1$ and $\varphi=0$. (a) Total FI with respect to the two-source separation $s$ and coherence $\gamma$  of the state $\ket{\Psi}$. (b) Total FI versus the separation $s$ and the entanglement measure concurrence $C$. (c) Total Fisher information for a fixed separation $s/\sigma=0.3$ with respect to coherence (in red) and concurrence (in blue).}
\label{FI}
\end{figure}

Fig.~\ref{FI} (a) illustrates the behavior of Fisher information $F_{tot}$ with respect to both $s$ and $C$. It is noticed that within the region $0\le s\lesssim 3\sigma$, the FI increases as the amount of entanglement $C$ increases for any fixed separation $s$. This is a region of most interest as it includes all the superresolution domain when two-source separation is beyond the diffraction limit, i.e., $s\in [0, \sim 1.22\sigma)$. Therefore, one is led to the conclusion that entanglement is a crucial feature which can enhance the measurement precision  for superresolution. In the remaining region $s\gtrsim 3\sigma$, i.e., well within the conventional resolution capability, the FI $F_{tot}$ remains at a very high value for all entanglement values. This is sensible because the separation is much greater than the diffraction limit and even conventional method is able to reach high measurement precision. It is worth to mention that as the degree of entanglement is also dependent on the separation $s$, its maximally reachable value is a function of $s$.

To compare the entanglement effect with that of coherence, we re-express the FI in terms of the degree of coherence $|\gamma|$, 
\beqa
\label{Ftot}
F_{tot}=\frac{1}{4\sigma^{2}}-\frac{|\gamma| d(4\sigma^{2}-s^{2})}{16\sigma^{4}} -\frac{|\gamma|^{2} d^{2}s^{2}}{8(1+|\gamma|^{2}+2d|\gamma|)}.
\eeqa
Fig.~\ref{FI} (b) illustrates its behavior with respect to both $s$ and $C$. Apparently, within the region of interest $0\le s\lesssim 3\sigma$, FI is behaving oppositely to that of entanglement. That is, the increase of degree of coherence will decrease the Fisher information (or measurement precision). Particularly, when it reaches maximum coherence $|\gamma|=1$, the Fisher information $F_{tot}$ completely vanishes. Therefore, coherence is an unwanted feature for superresolution , consistent with results from previous studies \cite{Larson2018O,Hradil2019O}. 

For a better comparison, Fig.~\ref{FI} (c) illustrates the properties of FI for a fixed separation $s=0.3 \sigma$, where the decreasing and increasing behaviors of $F_{tot}$ with respect to both entanglement $C$ and coherence $\gamma$ are shown explicitly. Therefore, entanglement is clearly the favorable feature over coherence for superresolution.


\section{Two parameter estimation}
 
While single-parameter estimation is relatively straightforward to handle, there are often more than just one parameter being unknown and need to be measured in practice. A multi-parameter estimation analysis needs to be applied \cite{LiuJPA2020}. In this section, we consider the two-parameter estimation when an additional parameter (relevant to either entanglement or coherence) is also unknown besides the two-source separation $s$.

We first analyze the case when both separation $s$ and entanglement $C$ are unknown. As shown in section 2, entanglement $C$ depends both on the separation $s$ and the parameter $\theta$. Therefore, for convenience, we analyze the Fisher information by estimating both $s$ and $\theta$. Similar to the single-parameter estimation case, the measurement precision quantities (or Fisher information) $H_{s}$ and $H_{\theta}$ for the two parameter case also correspond to the Cram\'er-Rao lower bound, i.e., $Var(s)\geq H^{-1}_{s}$ and $Var(\theta)\geq H^{-1}_{\theta}$, for the respective parameters $s$ and $\theta$. The difference is that $H_{s}$ and $H_{\theta}$ are now related to a $2\times2$ quantum Fisher information matrix (QFIM) \cite{Larson2018O, LiuJPA2020, Rehacek2017PRA} $\hat{F}$,
\beqa
\label{Hquantity}
H_{s}=F_{ss}- \frac{F^{2}_{s\theta}}{F_{\theta\theta}}, \quad \text{and} \quad H_{\theta}=F_{\theta\theta}- \frac{F^{2}_{s\theta}}{F_{ss}},
\eeqa
where $F_{ij}$ with $i,j=s,\theta$ are the matrix elements that can be achieved as
\begin{eqnarray}
F_{ij}=\frac{1}{2}Tr[(L_{i}L_{j}+L_{j}L_{i})\rho_s].
\end{eqnarray}
Here $\rho_s$ is the state of the relevant (spatial) degree of freedom that is obtained as $\rho_{s}=Tr_{\{\ket{\phi_1},\ket{\phi_1^{\perp}}\}}\left[|\Psi\ra\la\Psi|\right]$, and $L_{i(j)}$ are the symmetric logarithmic derivative (SLD) operators with respect to the parameter $i (j)$ and are defined as 
\beq
L_{i}=\sum_{k,l=1}^{4}\frac{2}{\lambda_{k}+\lambda_{l}}(\bra{e_{k}}\frac{\partial\rho_s}{\partial i}\ket{e_{l}})\ket{e_{k}}\bra{e_{l}}.
\eeq
Note that the commutator of operators $L_{i}$ satisfy the zero-expectation value relation $Tr[\rho_{s}[L_{i},L_{j}]]=0$, indicating that a single measurement can be optimal and simultaneously obtained for both $s$ and $\theta$. The vectors $|e_k\ra$ with $k=1,2$ are the eigenvectors of $\rho_{s}$ given as $\ket{e_{1}}=(\ket{h_{-}}-\ket{h_{+}})/\sqrt{2(1-d)}$ and $\ket{e_{2}}=(\ket{h_{-}}+\ket{h_{+}})/\sqrt{2(1+d)}$ with corresponding eigenvalues $\lambda_{1}=(d-1)(\cos\theta-1)/2(1+d\cos\theta)$ and $\lambda_{2}=(d+1)(\cos\theta+1)/2(1+d\cos\theta)$. The vectors $|e_k\ra$ with $k=3,4$ are transformed from $|e_{1,2}\ra$, i.e., $\ket{e_{3}}=a^{-1}_{3}\partial\ket{e_{1}}/\partial s$ and $\ket{e_{4}}=a^{-1}_{4}\partial\ket{e_{2}}/\partial s$, where the coefficients are given as 
\beqa
a_{3}&=&\sqrt{\frac{1-A}{16\sigma^{4}(1-d)}-\frac{B^{2}}{4(1-d)^{2}}},\notag \\
a_{4}&=&\sqrt{\frac{1+A}{(1+d)}-\frac{B^{2}}{4(1+d)^{2}}}. \notag
\eeqa
with $A=d(s^{2}/4-1)$ and $B=-ds/4\sigma^2$. Then the derivative of the density matrix $\rho_{s}$ can be expressed as
\beqa
\frac{\partial\rho_{s}}{\partial s}&=&\frac{B(1-\cos^{2}\theta)}{2(1+d\cos\theta)^{2}}[-\ket{e_{1}}\bra{e_{1}}+\ket{e_{2}}\bra{e_{2}}]\nonumber\\
&&+\lambda_{1}a_{3}[\ket{e_{3}}\bra{e_{1}}+\ket{e_{1}}\bra{e_{3}}]\nonumber\\
&&+\lambda_{2}a_{4}[\ket{e_{4}}\bra{e_{2}}+\ket{e_{2}}\bra{e_{4}}],\\
\frac{\partial\rho_{s}}{\partial \theta}&=&\frac{(1-d^{2})\sin\theta}{2(1+d\cos\theta)^{2}}[\ket{e_{1}}\bra{e_{1}}-\ket{e_{2}}\bra{e_{2}}].
\eeqa

As the four vectors $|e_k\ra$, $k=1,2,3,4$ form a new orthonormal basis \cite{Larson2018O}, the SLD operators represent various $4\times4$ matrices. Then the QFIM elements $F_{ij}$ can be specifically achieved through SLD operator matrix elements and expressed as 
\beqa
F_{ss}&=&\lambda_{1}([L_{s}]^{2}_{11}+[L_{s}]^{2}_{13})+\lambda_{2}([L_{s}]^{2}_{22}+[L_{s}]^{2}_{24}),\\
F_{\theta\theta}&=&\lambda_{1}[L_{\theta}]^{2}_{11}+\lambda_{2}[L_{\theta}]^{2}_{22}, \\
F_{s\theta}&=&\lambda_{1}[L_{s}]_{11}[L_{\theta}]_{11}+\lambda_{2}[L_{s}]_{22}[L_{\theta}]_{22},
\eeqa
where the relevant $L_i$ operator matrix elements are computed as 
\beqa
[L_{s}]_{11}=\frac{-B(1-\cos^{2}\theta)}{2\lambda_{1}(1+d\cos\theta)^{2}},\quad \left[L_s\right]_{13}=[L_{s}]_{31}=2a_{3},\\
\left[L_s\right]_{22}=\frac{-B(1-\cos^{2}\theta)}{2\lambda_{2}(1+d\cos\theta)^{2}},\quad \left[L_s\right]_{24}=[L_{s}]_{42}=2a_{4},\\
\left[L_{\theta}\right]_{11}=\frac{(1-d^{2})\sin\theta}{2\lambda_{1}(1+d\cos\theta)^{2}}, \quad \left[L_{\theta}\right]_{22}=\frac{-(1-d^{2})\sin\theta}{2\lambda_2(1+d\cos\theta)^{2}}.
\eeqa

\begin{figure}[t!]
\includegraphics[width=6.5cm]{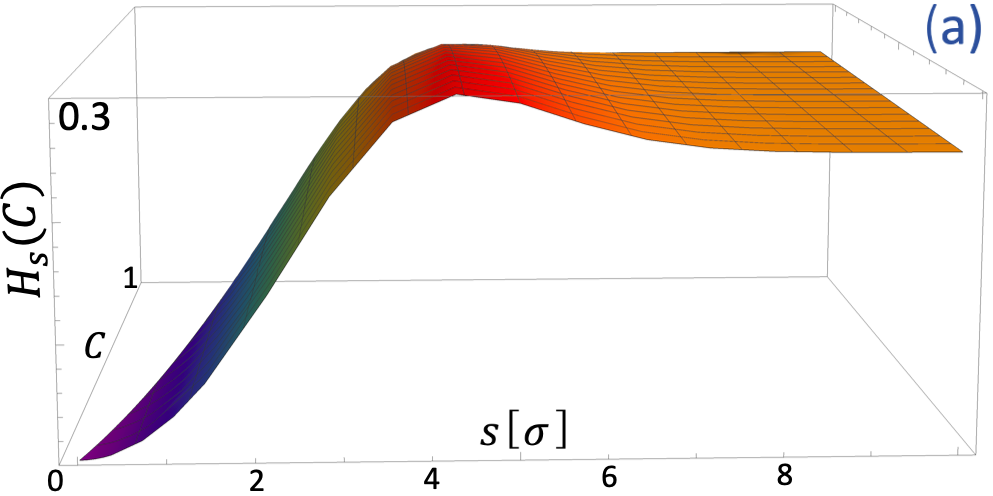}
\includegraphics[width=6.7cm]{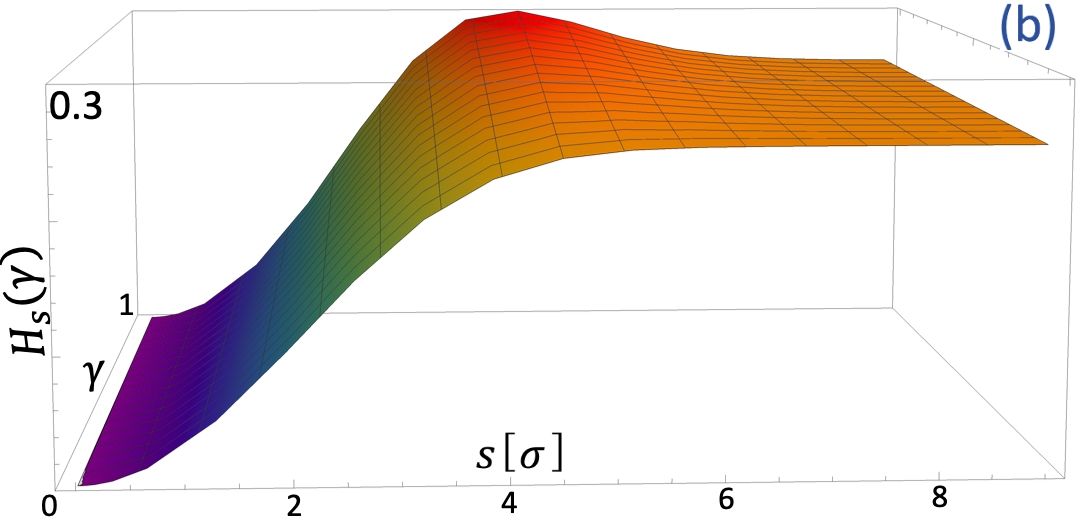}
\caption{(a) Both $s$ and the entanglement are unknown parameters. Precision $H_{s}$ with respect to the two-source separation $s$ and coherence $\gamma$ for $\sigma=1$ and $\varphi=0$.(b) Both $s$ and the coherence are unknown parameters for $\sigma=1$ and $\varphi=0$  Precision $H_{s}$ with respect to the two-source separation $s$ and coherence $\gamma$.}
\label{FImulti}
\end{figure}

Then the Fisher information quantities $H_{s}$ and $H_{\theta}$ are fully determined, through equation (\ref{Hquantity}), as functions of  $s$ and $\theta$ (entanglement $C$). The behavior of Fisher information $H_{s}=H_{s}(s,C)$ with respect to the estimation of separation $s$ is illustrated in Fig.~\ref{FImulti} (a). It is noted that the Fisher information disappears, i.e., $H_{s}(s,C)=0$, only in the zero-entanglement case, $C=0$, where the two-source separation also disappears $s=0$. This simply means that any nonzero entanglement $C\neq0$ will ensure a non-vanishing Fisher information $H_{s}(s,C)>0$. This echoes with the conclusion from the single-parameter estimation case that entanglement is indeed a favorable feature for superresolution. 

To compare with the effect coherence, we also carry out the multi-parameter estimation analysis for the two-source field (\ref{ent}) when both the separation $s$ and the degree of coherence $|\gamma|$ are unknown. Then the precision quantities (Fisher information) $H_{s}(s, \gamma)$ and $H_{\gamma}(s, \gamma)$ are obtained similarly through
\beqa
\label{Hgamma}
H_{s}(s,\gamma)=G_{ss}- \frac{G^{2}_{s\gamma}}{G_{\gamma\gamma}}, \quad \text{and} \quad H_{\gamma}(s,\gamma)=G_{\gamma\gamma}- \frac{G^{2}_{s\gamma}}{G_{ss}},
\eeqa
where $G_{ij}$ with $i,j=s,\gamma$ are the QFIM matrix elements given as $G_{ij}=\frac{1}{2}Tr[(L_{i}L_{j}+L_{j}L_{i})\rho_s]$ with $i,j=s,\gamma$ and $L_{j}$ being the corresponding SLD operators. Fig.~\ref{FImulti} (b) illustrates the behavior of precision quantity $H_{s}(s, \gamma)$ with respect to the two-source separation $s$ and coherence $|\gamma|$. Consistent with previous literature \cite{Larson2018O}, when the separation vanishes $s=0$, all values of coherence are leading to the disappearance of Fisher information $H_{s}(s, \gamma)$. This simply means that coherence can't help to improve measurement precision. 

Therefore, for the two-parameter estimation case, entanglement (in contrast to coherence) is again shown to be the favorable feature preferred by superresolution.


 \section{Conclusion}
In summary, we have investigated superresolution of two equal-intensity point sources with arbitrary non-controllable coherence and entanglement. The effects of the degree of entanglement and coherence on the measurement precision are analyzed systematically for both single- and two-parameter estimations. When the two-source separation is the only unknown parameter, it is found that while coherence is harmful, entanglement on the contrary can enhance the quality of superresolution significantly, i.e., higher degree of entanglement will lead to higher measurement precision of superresolution. When an additional parameter (either coherence or entanglement) is also taken to be unknown, entanglement is again shown to be beneficial to the measurement precision while coherence can't help at all.  

Our result suggests that superresolution measurements should include entanglement at two different levels. First, it is ideal to gain {\em a priori} knowledge of the amount of entanglement (corresponding to the single-parameter estimation case), as it will enhance the separation measurement precision significantly. Second, if not feasible to pre-know the amount entanglement (corresponding to the two-parameter estimation case), it is still possible to get high quality superresolution if the amount of entanglement can be increased as much as possible. \\


\noindent{\bf Acknowledgments.}  We acknowledge partial supports from U.S. Army under contact No. W15QKN-18-D-0040 and from Stevens Institute of Technology.\\



{}

\end{document}